\documentclass[
               twocolumn,
               prd,
	       nofootinbib,
               showpacs,preprintnumbers,
               amsmath,amssymb]
              {revtex4}
\usepackage{graphicx}  
\usepackage{amssymb}
\usepackage{amsmath}
\allowdisplaybreaks[1]
\def\e{{\rm e}}

\def\d{\partial}
\def\l{\left(}
\def\r{\right)}

\newcommand{\be}{\begin{equation}}
\newcommand{\ee}{\end{equation}}
\newcommand{\bea}{\begin{eqnarray}}
\newcommand{\eea}{\end{eqnarray}}
\newcommand{\bg}{\begin{gather}}
\newcommand{\eg}{\end{gather}}
\newcommand{\bseq}{\begin{subequations}}
\newcommand{\eseq}{\end{subequations}}

\begin{document} 

\title{More on ghosts in DGP model} 
\author{Dmitry Gorbunov$^a$, Kazuya Koyama$^b$ and Sergei Sibiryakov$^a$} 
\affiliation{%
  $^a$ Institute for Nuclear Research of the Russian Academy of Sciences,\\
  60th October Anniversary prospect 7a, Moscow 117312, Russia\\
%\vspace*{0.2cm} 
$^b$ Institute of Cosmology and 
Gravitation, University of Portsmouth, Portsmouth~PO1~2EG, UK 
\vspace*{0.3cm}
} 
%\date{15/08/05} 
\begin{abstract}
It is shown by an explicit calculation that 
the excitations about the
self-accelerating cosmological solution of the
Dvali--Gabadaze--Porrati model contain a ghost mode. This raises
serious doubts about viability of this solution. 
Our analysis reveals the similarity between the quadratic theory for the
perturbations around the self-accelerating Universe and an Abelian
gauge model with two St\"uckelberg fields. 
 
\end{abstract} 
\maketitle 

%04.20.Fy 	Canonical formalism, Lagrangians, and variational
%principles
%04.90.+e 	Other topics in general relativity and gravitation
%              (restricted to new topics in section 04)
%04.60.Ds Canonical quantization
%11.10.Ef 	Lagrangian and Hamiltonian approach

\section{Introduction and Summary}
Accelerated expansion of the Universe \cite{Riess:1998cb} 
is one of the most important
discoveries in cosmology. Usually, it is explained by the existence 
of tiny cosmological constant or some scalar fields in the framework of 
the Einstein's general relativity. However, it is worth exploring
an alternative possibility that general relativity breaks down on
cosmological scales  
and the accelerated expansion of the Universe results from
modification of gravitational laws themselves. 
Many attempts have been made in this direction, but, it has been recognized 
that it is extremely difficult to construct a consistent theory.
One of the problems often encountered by the attempts to modify gravity in
the infrared is appearance of ghost modes in the spectrum of the
theory. These are modes with the wrong sign in front of the kinetic
term. They have negative kinetic energy and lead to vacuum instability  
with respect to creation of ghost particles together with ordinary
matter particles, the interaction between ordinary matter and ghost
being mediated (at least) by gravity. 
The rate of the development of such
instability diverges due to infinite phase volume 
unless one introduces a Lorentz-violating cutoff, see e.g.
\cite{Cline:2003gs}.
 
An interesting model incorporating modification of gravitational laws
at large distances was proposed by Dvali, Gabadadze and Porrati (DGP)
in \cite{Dvali:2000hr}. The model
describes a brane with four-dimensional worldvolume, embedded into
flat 
five-dimensional bulk. Ordinary matter is supposed to be localized
on the brane, while gravity can propagate in the bulk. 
A crucial ingredient of the model is the induced Einstein-Hilbert
action on the brane. 
The action of the model is given by
\be
\label{act}
\begin{split}
S=\frac{1}{2\varkappa^2}\int d^5x\sqrt{-g}R
+\frac{1}{2\varkappa_4^2}\int d^4x\sqrt{-\gamma}~^{(4)}R
\\+
\int d^4x\sqrt{-\gamma}{\cal L}_{mat}\;,
\end{split}
\ee
where $R$ is the five-dimensional Ricci curvature, $^{(4)}R$ is the
intrinsic curvature of the brane computed using the induced metric
$\gamma_{\mu\nu}$, and ${\cal L}_{mat}$ is the Lagrangian of the
matter on the brane. 
Throughout the paper we use $(-,+,+,\ldots)$ convention for the 
signature of
the metric.
The induced gravity term is responsible for the
recovery of four-dimensional Einstein gravity at moderate 
scales\footnote{Let us mention that the mechanism of 
this recovery of four-dimensional
  gravity is rather non-trivial \cite{Deffayet:2001uk,Nicolis:2004qq}.},
while at distances larger than
\be
\label{r_c}
r_c\equiv\frac{\varkappa^2}{2\varkappa_4^2}\;
\ee 
gravity is five-dimensional. Cosmology in the DGP model is governed by
the following modification of the Friedman equation 
\cite{Deffayet:2001pu},
\be
\label{Friedman}
H^2=\frac{\varkappa_4^2\rho_{mat}}{3}\mp\frac{H}{r_c}\;,
\ee
where $H$ is the Hubble parameter and $\rho_{mat}$ is the matter density on
the brane. Two possible choices of sign in (\ref{Friedman}) give two
branches of the cosmological evolution. The upper sign corresponds to the
Universe whose expansion, 
in the absence of the cosmological constant on the
brane, decelerates at late times, the Hubble parameter tending
to zero as the matter on the brane dissolves. 
We call this branch of solutions the Friedman--Robertson--Walker (FRW)
branch. 
On the other hand, the
choice of the lower sign in (\ref{Friedman}) makes possible de Sitter
expansion of the Universe with the Hubble parameter $H=1/r_c$ 
even in the absence of matter.  
Thus, the latter branch contains the self-accelerating solution, where
the accelerated expansion of the Universe is realized without
introduction of the cosmological constant on the brane. Below, we will
refer to this branch of solutions as the self-accelerating branch.

However, the self-accelerating branch of solutions turns out to be
plagued by the ghost instability\footnote{This is not the case for the
FRW branch.}. This was first demonstrated in
\cite{Luty:2003vm,Nicolis:2004qq} using the boundary effective action
formalism. In \cite{Koyama:2005tx} this result was confirmed by
explicit calculation of the spectrum of linear perturbations in the
five-dimensional framework. The analysis was performed in the case when
the role of matter on the brane is played by non-zero brane tension
$\sigma\neq 0$. At $\sigma<0$ the ghost mode was identified with the
scalar field describing the brane bending, while at
$\sigma>0$ it was found that the ghost degree of freedom 
coincides with the helicity-0
component of the graviton, which turns out to be massive. 
By
continuity, one expects the ghost to be present also in the case
$\sigma=0$ which corresponds precisely to the self-accelerating
cosmological evolution. 

The purpose of this paper is to show the presence of ghost in the
self-accelerating Universe with $\sigma=0$ by an explicit
calculation. The analysis is subtler in this case than for $\sigma\neq
0$.  The reason is that at $\sigma=0$ the masses of the graviton and
the brane bending mode coincide and the two modes mix.  As a
consequence, it is impossible to diagonalize the quadratic Lagrangian
for these modes 
and single out the
Lagrangian for the ghost mode.  Instead, to demonstrate the existence
of ghost, we use the Hamiltonian approach
\cite{Abbott:1981ff,Deser:2001wx}.  We construct the Hamiltonian for
the helicity-0 excitations and find that it is unbounded from
below. For modes of high momentum, $k\gg H$, the mixing terms in the
Hamiltonian can be neglected and it decomposes into a sum of
Hamiltonians for a positive energy mode and ghost.

The paper is organized as follows. We start by considering in 
Sec.~\ref{abelian}  a simple model illustrating the subtleties
encountered by the analysis of the spectrum in the self-accelerating
Universe. In Sec.~\ref{perturbations} we find the perturbations about
the self-accelerating solution of the DGP model and construct the
four-dimensional effective action for the discrete graviton mode and
the brane-bending mode. In Sec.~\ref{hamiltonian} we
construct the Hamiltonian for the helicity-0 excitations and show
that it contains a ghost mode. Technical details of the calculation of
the Hamiltonian are presented in Appendix.

\section{Abelian example}
\label{abelian}

To set the stage for the analysis of the quadratic theory for
perturbations about the self-accelerating solution, let us consider a
simple toy model whose spectrum exhibits analogous properties.  We
consider an Abelian gauge theory coupled to two St\"uckelberg
fields\footnote{We thank R.~Rattazzi for suggesting this
model to us.}. The Lagrangian of the model has the form, \be {\cal
L}=-\frac{1}{4}F_{\mu\nu}^2-
\frac{f^2_1}{2}(\d_{\mu}\varphi_1+eA_{\mu})^2
+\frac{f_2^2}{2}(\d_{\mu}\varphi_2+eA_{\mu})^2\;,
\label{lagr}
\ee   
where $f^2_1, f^2_2 >0$. When the interaction between the vector field 
and the scalar bosons is switched off by setting $e=0$, the fields
$\varphi_1$ and 
$\varphi_2$ are an ordinary scalar field and ghost, respectively.
Below we assume $e\neq 0$. The system possesses gauge symmetry 
\be
\label{emtrans}
A_{\mu}\mapsto A_{\mu}-\frac{1}{e}\d_{\mu}\alpha~,~~~
\varphi_1\mapsto \varphi_1+\alpha~,~~~
\varphi_2\mapsto \varphi_2+\alpha\;,
\ee
which can be used to set $\varphi_2=0$. 
In the case 
$f_1^2\neq f_2^2$ one can diagonalize the Lagrangian (\ref{lagr})
by introducing the vector field
\be
\label{bmu}
B_{\mu}=A_{\mu}+\frac{f_1^2\d_{\mu}\varphi_1}{(f_1^2-f_2^2)e}\;.
\ee
The result of the diagonalization has the form
\be
\label{lagr1}
{\cal L}=-\frac{1}{4}B_{\mu\nu}^2-\frac{(f_1^2-f_2^2)e^2}{2}B^2_{\mu}
+\frac{f_1^2f_2^2}{2(f_1^2-f_2^2)}(\d_{\mu}\varphi_1)^2\;.  
\ee
It describes a massless scalar field and a massive vector with the
mass $m^2_B=(f_1^2-f_2^2)e^2$. If $f_1^2-f_2^2>0$ the scalar is 
ghost, while the vector field has positive mass squared. 
On the other hand, when
$f_1^2-f_2^2<0$ the sign in front of the scalar kinetic term is
correct, while the mass squared of the vector becomes negative. This
implies 
that the
longitudinal component of the vector is ghost in this case.

At the point $f_1^2-f_2^2=0$ in the parameter space the vector field
becomes massless. Simultaneously, 
the 
transformation (\ref{bmu}) becomes singular suggesting that 
the Lagrangian cannot be
diagonalized. In terms of fields $A_\mu$, $\varphi_1$
the Lagrangian reads
\be
\label{veclagr}
{\cal L}=-\frac{1}{4}F_{\mu\nu}^2-\frac{f_1^2}{2}(\d_\mu\varphi_1)^2
-f_1^2eA_{\mu}\d^\mu\varphi_1\;.
\ee 
The last term describes mixing between the vector and the scalar
fields. Note that, if this last term were absent,
the Lagrangian (\ref{veclagr}) would possess an Abelian gauge
symmetry which would make the helicity-0 component of the
vector unphysical, and  the theory would be ghost-free. 
However, the Lagrangian (\ref{veclagr}) as it stands, is not
gauge invariant and the helicity-0 component of the vector field
becomes a physical ghost through its mixing with the scalar field.

To demonstrate this we consider the Hamiltonian of the theory
(\ref{veclagr}). One concentrates on the
helicity-0 sector as the transverse degrees of freedom are unaffected
by the presence of the scalar field, and hence, their energy is
positive definite.
One performs the Fourier
decomposition
\be
\begin{split}
&A_0=\int\frac{d^3 k}{(2\pi)^{3/2}}\e^{i{\bf kx}}n({\bf k},t)\;,\\
&A_i=-i\int\frac{d^3 k}{(2\pi)^{3/2}}\e^{i{\bf kx}}
\frac{k_i}{|k|}\alpha({\bf k},t)\;,\\
&\varphi_1=\int\frac{d^3 k}{(2\pi)^{3/2}}\e^{i{\bf kx}}\phi({\bf
  k},t)\;.
\end{split}
\ee
Inserting these expressions into (\ref{veclagr}) and integrating out
the variable $n$ one obtains the Lagrangian for the fields $\alpha$,
$\phi$,
\be
\label{vlagr2}
\begin{split}
L_0=\int d^3k\bigg\{\frac{f_1^2}{2}
\left(1-\frac{(f_1e)^2}{k^2}\right)|\dot\phi|^2
-\frac{f_1^2e}{|k|}\dot\phi\dot\alpha^*\\
-\frac{f_1^2k^2}{2}|\phi|^2
+f_1^2e|k|\alpha\phi^*\bigg\}\;,
\end{split}
\ee
where dot denotes derivative with respect to time. 
In writing (\ref{vlagr2}) we made use of the relations 
$\alpha(-{\bf k},t)= \alpha^*({\bf k},t)$, etc. Introducing the
canonical momenta
\[
\pi_\alpha=-\frac{f_1^2e}{|k|}\dot\phi~,~~~~
\pi_\phi=-\frac{f_1^2e}{|k|}\dot\alpha
+f_1^2\left(1-\frac{(f_1e)^2}{k^2}\right)\dot\phi\;,
\] 
we obtain the Hamiltonian,
\be
\label{vham1a}
\begin{split}
{\cal H}_0=\int d^3k\bigg\{
-\frac{k}{f_1^2e}\pi_\phi\pi_\alpha^*
+\frac{-k^2+(f_1e)^2}{2f_1^2e^2}
|\pi_\alpha|^2\\
+\frac{f_1^2k^2}{2}|\phi|^2
-f_1^2ek\alpha\phi^*
\bigg\}\;.
\end{split}
\ee
After the canonical transformation
\be
\begin{split}
&\tilde\alpha=f_1e\alpha~,
~~~\tilde\phi=-f_1e\alpha+f_1k\phi~,~~~\\
&\pi_\alpha
=f_1e\l\tilde\pi_\alpha-\tilde\pi_\phi\r~,~~~
\pi_\phi=f_1k\tilde\pi_\phi\;,
\end{split}
\ee
it takes the following form 
\be
\label{vham1}
\begin{split}
{\cal H}_0=\int& d^3k\bigg\{
\frac{k^2+(f_1e)^2}{2}
|\tilde\pi_\phi|^2
-(f_1e)^2\tilde\pi_\phi\tilde\pi_\alpha^*\\
&+\frac{-k^2+(f_1e)^2}{2}
|\tilde\pi_\alpha|^2
+\frac{|\tilde\phi|^2}{2}-\frac{|\tilde\alpha|^2}{2}\bigg\}\;.
\end{split}
\ee
Clearly, this Hamiltonian is unbounded from below, negative energies
being associated with the field $\tilde\alpha$. For the modes of 
high spatial momentum, $k^2\gg (f_1e)^2$, one can neglect the mixing
terms in (\ref{vham1}); then, it is clear that the
mode $\tilde\alpha$ is ghost.

At generic values of the momentum $k$, 
the Hamiltonian (\ref{vham1}) 
represents one of the normal forms of
quadratic Hamiltonians. It 
cannot be diagonalized  by a
canonical transformation (see, e.g.  \cite{Arnold}).
This
fact is a manifestation of the {\it resonance } between the modes
$\tilde\alpha$ and $\tilde\phi$.
The solution of the equations of motion
following from (\ref{vham1}) has the form
\[
%\be
%\label{sol}
\begin{pmatrix}
\tilde\alpha\\
\tilde\phi
\end{pmatrix}
=a^{\pm}\!\!\begin{pmatrix}
1\\
-1
\end{pmatrix}\!\e^{\pm i |k|t}
+b^{\pm}\!\!\left[\!
\begin{pmatrix}
1\\
1
\end{pmatrix}
\!\mp\! i\frac{(f_1e)^2}{|k|}t\!
\begin{pmatrix}
1\\
-1
\end{pmatrix}\!\right]\!\e^{\pm i |k|t}
%\ee
\]
It contains a linearly growing part due to the 
resonance. Again, this part can be neglected
for high frequency modes and at
short time scales $t\sim\frac{1}{|k|}\ll \frac{|k|}{(f_1e)^2}$. The
solution in this regime becomes the sum of two purely oscillatory
modes.

\section{Perturbations in self-accelerating Universe}
\label{perturbations}

We now proceed to apply the Hamiltonian analysis, analogous to that
of the previous section, to the DGP model. In this section we study
the linear perturbations about the self-accelerating cosmological
solution, and construct quadratic four-dimensional action for the
localized modes.

The five dimensional background metric corresponding to the
self-accelerated branch has the form \cite{Deffayet:2001pu},
 \be
ds^2=dy^2+N^2(y)\gamma_{\mu\nu}(x)dx^{\mu}dx^{\nu}\;,
\ee
where $N(y)=1+H|y|$, and $\gamma_{\mu\nu}$ is the four-dimensional de
Sitter (dS) metric with the Hubble parameter $H$. The brane is located at
$y=0$, and $Z_2$ symmetry across the brane is imposed.
The case of vanishing brane tension, which is of primary interest to
us, corresponds to  
\[
%\label{self}
H=\frac{1}{r_c}\;.
\]
However, we will not use this relation for somewhile in order to
be able to compare the cases of vanishing and non-zero
brane tensions. 

Let us now consider perturbations of the metric. We impose the gauge 
\be
\label{gauge}
\delta g_{55}=\delta g_{5\mu}=0, \ee and write
$g_{\mu\nu}=N^2\gamma_{\mu\nu}+h_{\mu\nu}(x,y)$.  For perturbations
obeying equations of motion it is possible to impose in addition to
(\ref{gauge}) the transverse-traceless (TT) gauge $h_\mu^\mu=0$,
$\nabla_\mu h^\mu_\nu=0$, where $\nabla_\mu$ denotes covariant
derivative with respect to the metric $\gamma_{\mu\nu}$, and
indices are raised using the metric $\gamma^{\mu\nu}$. In this gauge the
Einstein's equations reduce to
\begin{equation}
\label{graviton-in-bulk}
h_{\mu\nu}''
+\frac{1}{N^2}\l \Box h_{\mu\nu} 
-4H^2 h_{\mu\nu}\r=0\;,
\end{equation}
where $\Box\equiv\nabla_{\lambda}\nabla^{\lambda}$.
We will call the coordinate frame where the metric is TT the bulk
frame. 
In this frame the brane is displaced from the origin. We parametrize
the perturbation of its position by $y=\varphi(x)$. 
With the account for the brane bending one obtains the following 
junction condition at the brane in the TT gauge 
\cite{Koyama:2005tx},
\be
\begin{split}
\label{h-boundary}
\frac{1}{2}h_{\mu\nu}'+&\frac{r_c}{2}\Box h_{\mu\nu}
-H(1+Hr_c)h_{\mu\nu}\\
&=
(2Hr_c-1)(\nabla_\mu\nabla_\nu +H^2\gamma_{\mu\nu})\varphi\;.
\end{split}
\ee
Taking the trace of this equation results in the equation of motion
for the field $\varphi$,
\begin{equation}
\label{bending-equation}
\Box\varphi+4H^2\varphi=0\;.
\end{equation} 
Let us summarize the results about solutions of Eqs.~(\ref{graviton-in-bulk}), 
(\ref{h-boundary}) in the case $Hr_c\neq 1$ \cite{Koyama:2005tx}.
First, there are solutions leaving the brane at rest,
$\varphi=0$. They describe a tower of modes of the form
$h_{\mu\nu}(x,y)=\chi^{(m)}_{\mu\nu}(x)F_m(y)$. The fields 
$\chi^{(m)}_{\mu\nu}$
satisfy the four-dimensional equation 
$\Box \chi^{(m)}_{\mu\nu}=(m^2+2H^2)\chi^{(m)}_{\mu\nu}$
for massive spin-2 fields in dS space-time. The
functions $F_m(y)$ obey the following equations,
\begin{align}
\label{m-bulk}
&F''_m+\frac{m^2-2H^2}{N^2}F_m=0\;,~~~~y>0\;,\\
\label{m-boundary}
&F'_m-2HF_m=-m^2r_cF_m\;,~~~~y=0,
\end{align}
where prime denotes differentiation with respect to $y$.  
There is a continuum spectrum of modes with masses 
$m^2\geq (9/4)H^2$, and a normalizable mode 
\be
\label{dismode}
F_{m_d}(y)=[N(y)]^{-1+\frac{1}{Hr_c}}
\ee
with the mass
\be
\label{m_d}
m^2_d=\frac{3Hr_c-1}{r_c^2}\;.
\ee
Second, there is a discrete mode with non-zero brane bending.
The corresponding perturbation of the metric has the form,
\be
\label{scal}
h_{\mu\nu}(x,y)=\frac{1-2Hr_c}{H(1-Hr_c)}
(\nabla_\mu\nabla_\nu+H^2\gamma_{\mu\nu})\varphi(x)\;.
\ee
This is a solution with $m^2=2H^2$. In \cite{Koyama:2005tx} the
effective quadratic action for the two discrete modes was
constructed for the case $Hr_c\neq 1$. As the masses of
the modes are different in this case, 
$m_d^2\neq 2H^2$, the Lagrangian decomposes into
the Lagrangians for the massive spin-2 field $\chi_{\mu\nu}$ and the scalar
field $\varphi$. It was found that at $Hr_c<1$ the scalar field is a
ghost, while the tensor field is well behaved. 
[It is worth mentioning that the region $Hr_c<1$ corresponds to the
unphysical case of negative brane tension. Nevertheless, considering
this regime does make sense if the value of $Hr_c$ is close enough to unity.]
On the other hand, 
at $Hr_c>1$ the 
ghost mode coincides with the helicity-0
component of the field $\chi_{\mu\nu}$, while the scalar field has the
correct sign in front of its kinetic term. At the point $Hr_c=1$ the
masses of the two
modes coincide, and the
solution (\ref{scal}) becomes singular.
This signals that at the point $Hr_c=1$ the Lagrangian for the
discrete modes cannot be diagonalized.  
One concludes that the 
situation is analogous to the situation in the Abelian model
of the previous section.

Let us concentrate on the sector of the localized modes.  Keeping
these modes only, the expression for metric perturbations at $Hr_c\neq
1$ reads 
\[
\begin{split}
h_{\mu\nu}(x,y)=&\chi^{(m_d)}_{\mu\nu}(x)F_{m_d}(y)\\
&+
\frac{1-2Hr_c}{H(1-Hr_c)}
(\nabla_\mu\nabla_\nu+H^2\gamma_{\mu\nu})\varphi(x)\;.
\end{split}
\]
To obtain a
non-singular Lagrangian for the localized modes 
in the limit $Hr_c\to 1$ corresponding to 
the
self-accelerating Universe we  
perform a field redefinition 
(cf. Eq.~(\ref{bmu}))
\[
%\be
%\label{fieldredef}
\chi^{(m_d)}_{\mu\nu}(x)=A_{\mu\nu}(x)-\frac{1-2Hr_c}{H(1-Hr_c)}
(\nabla_\mu\nabla_\nu+H^2\gamma_{\mu\nu})\varphi(x)\;.
%\ee
\]
In terms of the fields $A_{\mu\nu}$, $\varphi$ the 
KK decomposition of the metric
perturbation reads
\[
%\be
\begin{split}
%\label{KK}
&h_{\mu\nu}(x,y)=A_{\mu\nu}(x)F_{m_d}(y)\\
&+\frac{1-2Hr_c}{H(1-Hr_c)}
(\nabla_\mu\nabla_\nu+H^2\gamma_{\mu\nu})\varphi(x)
(1-F_{m_d}(y))\;.
\end{split}
%\ee
\]
Now, we can take the limit $Hr_c\to 1$, keeping both fields
$A_{\mu\nu}(x)$ and $\varphi(x)$ finite. Using Eq.~(\ref{dismode})
we obtain, 
\be
\label{hAB}
h_{\mu\nu}=A_{\mu\nu}(x)
+\frac{1}{H}(\nabla_\mu\nabla_\nu+H^2\gamma_{\mu\nu})\varphi(x) \log\l
1+Hy\r\;.  \ee We note in passing that this representation of
$h_{\mu\nu}$ in terms of localized modes can be obtained directly in
the case $Hr_c=1$ starting from the Einstein equations in the bulk and
junction conditions on the brane. 

From now on we set $Hr_c=1$. The field $\varphi$ still obeys
Eq.~\eqref{bending-equation}. To obtain the equations of motion for
the field $A_{\mu\nu}$ we plug the expression (\ref{hAB}) into
Eqs.~(\ref{graviton-in-bulk}), (\ref{h-boundary}). This yields 
\be
\label{AB}
\Box A_{\mu\nu}-4H^2 A_{\mu\nu}=H\l \nabla_\mu\nabla_\nu
+H^2\gamma_{\mu\nu}\r\varphi\;.  
\ee 
Note, that the TT condition for
the metric perturbation $h_{\mu\nu}$ implies that $A_{\mu\nu}$
must be also TT.

We proceed to 
construct the effective quadratic action for the fields 
$A_{\mu \nu}$
and $\varphi$.  
The full quadratic action has the form:
%\be
\begin{multline}
\label{quadract}
S^{(2)}=\frac{1}{4\varkappa^2}
\int d^5x\sqrt{-g}\delta g^{MN}\delta G^{(5)}_{MN}\\
+\frac{1}{4\varkappa^2_4}\int d^4x\sqrt{-\gamma}
\delta g^{\mu\nu}\delta G^{(4)}_{\mu\nu}\;,
\end{multline}
%\ee
where $\delta G^{(5)}_{MN}$, $\delta G^{(4)}_{\mu\nu}$ are the
variations of the five-dimensional and four-dimensional Einstein's
tensors, respectively.
We impose the gauge
(\ref{gauge})
and write the action (\ref{quadract})
in the Gaussian normal (GN) coordinate frame
where the brane is at rest,
\begin{widetext}
\be
\label{SGN}
\begin{split}
S^{(2)}=-\frac{1}{2\varkappa^2}\bigg[\int_{+0}^{+\infty}
dy\;d^4x\sqrt{-\gamma}\bigg(-\frac{1}{2}\bar{h}^{\mu\nu}\bar{h}''_{\mu\nu}
+\frac{1}{2}\bar{h}\,\bar{h}''
+\frac{1}{N^2}\bar{h}^{\mu\nu}X_{\mu\nu}(\bar{h})
+\frac{H^2}{N^2}\bar{h}^{\mu\nu}\bar{h}_{\mu\nu}
-\frac{H^2}{N^2}\bar{h}^2\bigg)\\
+\int d^4x\sqrt{-\gamma}\bigg(\frac{1}{H}\bar{h}^{\mu\nu}X_{\mu\nu}(\bar{h})
+H\bar{h}^{\mu\nu}\bar{h}_{\mu\nu}-H\bar{h}^2
-\frac{1}{2}\bar{h}^{\mu\nu}\bar{h}_{\mu\nu}'
+\frac{1}{2}\bar{h}\,\bar{h}'\bigg)\bigg]\;.
\end{split}
\ee
\end{widetext}
~~
\newpage
~~
\newpage
\noindent
Here $\bar h_{\mu\nu}$ is the perturbation of the metric in the GN
frame, $\bar h\equiv \bar h_{\mu}^\mu$ and
%\newpage
\[
%\be
\begin{split}
X_{\mu\nu}&(\bar{h})\equiv\delta G_{\mu\nu}^{(4)}+3H^2\bar{h}_{\mu\nu}\\
=
&-\frac{1}{2}\l \Box\bar{h}_{\mu\nu}-
\nabla_{\mu}\nabla_{\alpha}\bar{h}^{\alpha}_{\nu}-
\nabla_{\nu}\nabla_{\alpha}\bar{h}^{\alpha}_{\mu}
+\nabla_{\mu}\nabla_{\nu}\bar{h}\r \\
&-\frac{1}{2}\gamma_{\mu\nu}
\l \nabla_{\alpha}\nabla_{\beta}\bar{h}^{\alpha\beta}-\Box\bar{h}\r
+H^2\l \bar{h}_{\mu\nu}+\frac{1}{2}\gamma_{\mu\nu}\bar{h}\r\;.
\end{split}
%\ee
\]
%\end{widetext}
In deriving the expression (\ref{SGN}) we included
the contributions from the first integral in 
(\ref{quadract}) which are proportional to 
$\delta(y)$ into the integral over the brane.
Let us note that the action, restricted to the gauge (\ref{gauge})
does not give $(55)$ and $(5\mu)$ Einstein
equations. So, a priori, 
the latter equations must be imposed as constraints. However, we will
see that the equations of motion coming from the effective action
imply that the metric perturbations are TT. 
This, in turn, implies that the $(55)$ and $(5\mu)$ Einstein
equations are automatically satisfied. 

The metrics in GN and bulk frames are 
related by the 
gauge transformation
\be
\label{transf}
\bar h_{\mu\nu}=h_{\mu\nu}-
\frac{2N}{H}\nabla_\mu\nabla_\nu\varphi-2HN\gamma_{\mu\nu}\varphi\;.
\ee 
To obtain the effective
action for $A_{\mu\nu}$, $\varphi$ one takes $h_{\mu\nu}$ in the form
(\ref{hAB}) and plugs the expression (\ref{transf}) in (\ref{SGN}).
Note that in spite of the logarithmic growth of the metric
perturbation (\ref{hAB}) into the fifth dimension, the integrals over
$y$ in (\ref{SGN}) are finite due to the presence of $1/N^2(y)$ warp
factors.  After straightforward calculation one obtains
\begin{widetext}
\be
\label{effact}
\begin{split}
S_{eff}=\frac{1}{\varkappa^2H}\int d^4x\sqrt{-\gamma}
\bigg\{&-A^{\mu\nu}X_{\mu\nu}(A)-H^2A^{\mu\nu}A_{\mu\nu}
+H^2A^2\\
&-H(A^{\mu\nu}\nabla_\mu\nabla_\nu\varphi-A\Box\varphi-3H^2A\varphi)
-\frac{9H^2}{4}\varphi(\Box+4H^2)\varphi\bigg\}\;,
\end{split}
\ee
\end{widetext}
where we introduced the notation $A\equiv A_\mu^\mu$. 
Let us stress 
that it would be incorrect to impose the TT condition on the
field $A_{\mu\nu}$ in the action (\ref{effact}). In particular,
assuming $A_{\mu\nu}$ to be TT would lead
to the absence of the term describing mixing between
the fields $A_{\mu\nu}$ and $\varphi$ in (\ref{effact}), 
and hence, to incorrect
field equations. On the other hand, let us show that 
the equations of motion following from the action (\ref{effact}) 
entail both the TT conditions 
and the field equations (\ref{bending-equation}), (\ref{AB}). 
By varying the action (\ref{effact}) we obtain 
\begin{align}
&-2X_{\mu\nu}(A)-2H^2A_{\mu\nu}+2H^2\gamma_{\mu\nu}A\nonumber\\
&~~~~~~~~~-H\nabla_\mu\nabla_\nu\varphi+H\gamma_{\mu\nu}\Box\varphi
+3H^3\gamma_{\mu\nu}\varphi=0\;,
\label{effeq1}\\
\label{effeq2}
&-\nabla_\mu\nabla_\nu A^{\mu\nu}+\Box A+3H^2A
-\frac{9H}{2}(\Box+4H^2)\varphi=0\;.
\end{align}
Taking the covariant divergence and the trace of the first equation
one obtains
\begin{gather}
\label{divtr}
\nabla_\mu A^\mu_\nu-\nabla_\nu A=0\;,\\
\label{tr1}
2\nabla_\mu\nabla_\nu A^{\mu\nu}-2\Box A+
3H(\Box+4H^2)\varphi=0\;,
\end{gather}
where in deriving (\ref{divtr}) we made use of the identity 
$\nabla_\mu X^\mu_\nu(A)\equiv 0$. 
Equations \eqref{divtr}, \eqref{tr1} imply Eq.~(\ref{bending-equation}). 
Now, Eqs.~(\ref{effeq2}), (\ref{divtr}), (\ref{bending-equation}) yield the TT
conditions,
\be
\nabla_\mu A^\mu_\nu=0\;,~~~~A=0\;.
\ee
With the use of Eq.~(\ref{bending-equation}) and
the TT conditions equation 
(\ref{effeq1}) is reduced to Eq.~(\ref{AB}). 

Before proceeding further, let us make the following comment. 
The first line in (\ref{effact}) coincides with the
quadratic Lagrangian for the Fierz--Pauli theory 
of massive graviton with the mass $m^2=2H^2$ in dS 
background. Thus, it is invariant \cite{Deser:2001wx}
under gauge
transformations
\be
\label{DWgauge}
A_{\mu\nu}\mapsto A_{\mu\nu}
+(\nabla_\mu\nabla_\nu+H^2\gamma_{\mu\nu})\omega(x)\;.
\ee
The mixing of the graviton field $A_{\mu\nu}$ with the scalar $\varphi$
in (\ref{effact}) breaks this symmetry explicitly. Let us restore the 
symmetry by
introducing the St\"uckelberg field $\psi$.
Namely, let us consider
the following action,
\begin{widetext}
\be
\label{actinv}
\begin{split}
\tilde S_{eff}=\frac{1}{\varkappa^2H}\int d^4x\sqrt{-\gamma}
\bigg\{&-A^{\mu\nu}X_{\mu\nu}(A)-H^2A^{\mu\nu}A_{\mu\nu}
+H^2A^2\\
&+\frac{3}{4}\Big(\Big[(\nabla_\mu\nabla_\nu+H^2\gamma_{\mu\nu})\varphi
-\frac{2H}{3}A_{\mu\nu}\Big]^2-
\Big[(\Box+4H^2)\varphi-\frac{2H}{3}A\Big]^2\Big)\\
&-\frac{3}{4}\Big(\Big[(\nabla_\mu\nabla_\nu+H^2\gamma_{\mu\nu})\psi
-\frac{2H}{3}A_{\mu\nu}\Big]^2-
\Big[(\Box+4H^2)\psi-\frac{2H}{3}A\Big]^2\Big)\bigg\}\;.
\end{split}
\ee 
\end{widetext}
Note that the terms with four derivatives in   
(\ref{actinv})
cancel out after integration by parts.
The action (\ref{actinv}) is invariant under the gauge transformation
(\ref{DWgauge}) supplemented by 
\be
\label{gravtrans}
\varphi\mapsto\varphi+\frac{2H}{3}\omega~,~~~~
\psi\mapsto\psi+\frac{2H}{3}\omega\;.
\ee
It is straightforward to check that 
in the gauge $\psi=0$ the action (\ref{actinv}) reduces 
to (\ref{effact}). The form of the Lagrangian (\ref{actinv}) is
similar to the Lagrangian (\ref{lagr}) of the Abelian model of
Sec.~\ref{abelian}; it corresponds to equal coefficients in front of
the Lagrangians for the St\"uckelberg fields, $f_1^2=f_2^2$. 
In fact, one can check, that 
the quadratic
action for the perturbations about the self-accelerating Universe can
be written as an action for a massive graviton with the mass
$m^2=2H^2$ and two St\"uckelberg fields in dS space-time 
at any value of the parameter $Hr_c$. We will not pursue this issue
further in this paper.

\section{Hamiltonian for helicity-0 excitations}
\label{hamiltonian}
Let us come back to the four-dimensional theory with quadratic
action \eqref{effact}. 
Our aim now is to show that this theory has ghost 
in dS space-time. We adopt the approach of
Refs.~\cite{Abbott:1981ff,Deser:2001wx} and construct
the Hamiltonian for the helicity-0 excitations.

To simplify the formulas we set in this section the four-dimensional
Planck mass equal to one, $\varkappa^2H=2\varkappa_4^2=1$. 
We use the following explicit form of dS metric:
\be
\label{dSmetric}
ds^2=-dt^2+f^2(t)\delta_{ij}dx^i dx^j\;,
\ee  
where
\be
\label{ft}
f(t)=\e^{Ht}\;,
\ee
and $i$ runs from 1 to 3.
One decomposes various tensors into spatial and
time components, and introduces
\be
\label{dec}
A_{ij},~~~~~~N_i=A_{0i},~~~~~~n=-A_{00}/2\;.
\ee 
These notations are reminiscent of the standard notations in the
Hamiltonian approach to gravity. Indeed, 
the first term in
(\ref{effact}) represents the quadratic action for gravitational
perturbations in dS background with $\delta
g_{\mu\nu}=A_{\mu\nu}$. The quantities in Eqs.~(\ref{dec}) correspond
to 
linear
perturbations of the spatial metric, $g_{ij}=f^2\delta_{ij}+A_{ij}$,
the shift functions, $N_i=g_{0i}=A_{0i}$, and the lapse function,
$N=(-g^{00})^{-1/2}=1+n$.   
The next step is to perform the spatial Fourier decomposition. 
We are interested in the helicity-0 sector, as we expect the ghost
mode to be present there. (As to the helicity $\pm 2$ and $\pm 1$
sectors, they are not affected by the presence of the scalar $\varphi$
and do not contain instabilities \cite{Deser:2001wx}.) So, we write
\bseq
\label{Four}
\begin{align}
f^{-1/2}A_{ij}=
\int\frac{d^3k}{(2\pi)^{3/2}}\e^{i{\bf kx}}
\bigg\{&\left(\delta_{ij}-\frac{k_ik_j}{k^2}\right)\alpha_1({\bf k},t)
\notag\\
&+\frac{k_ik_j}{k^2}\alpha_2({\bf k},t)\bigg\}\;,
\label{ANFour}\\
f^{1/2}N_i=-i
\int\frac{d^3k}{(2\pi)^{3/2}}\e^{i{\bf kx}}
&\frac{k_i}{|k|}
\nu({\bf k},t)\;,\\
\label{nphiFour}
f^{3/2}n=
\int\frac{d^3k}{(2\pi)^{3/2}}\e^{i{\bf kx}}n({\bf k},&t)\;,\\
Hf^{3/2}\varphi=
\int\frac{d^3k}{(2\pi)^{3/2}}\e^{i{\bf kx}}\phi(&{\bf k},t)\;.
\end{align}
\eseq 
Note that we use the same notation $n$ for the Fourier transform of
the lapse 
function; this will not lead to confusion. 
The $f$ factors on the l.h.s of Eqs.~(\ref{Four}) are chosen in such a
way that the time-dependence of the resulting Hamiltonian is simplified, 
see below.

All but two degrees of freedom introduced in (\ref{Four}) are
non-dynamical and are eliminated by making use of the constraints. 
The details of this procedure are presented in Appendix.
Once this is
done 
the Hamiltonian is expressed in terms of two 
dynamical variables, the brane bending mode $\phi$ and the helicity-0
excitation of the graviton
$v \equiv \alpha_1 -\alpha_2$.
The Hamiltonian has the form,
\be
\label{Ham2}
\begin{split}
&{\cal H}_0=\int d^3k\bigg\{
\left(\frac{\hat k^4}{3H^4}
+\frac{\hat k^2}{H^2}+\frac{3}{2}\right)|\pi_v|^2
+\frac{2\hat k^2}{3H^2}\pi_v\pi_\phi^*\\
&-\left(\frac{\hat k^2}{3H}-\frac{H}{2}\right)\pi_v v^*
-\frac{\hat k^2}{2H}\pi_v\phi^*
-\left(\frac{\hat k^2}{3H}-\frac{5H}{2}\right)\pi_\phi\phi^*\\
&+\frac{\hat k^2}{6}v\phi^*-\frac{\hat k^4}{12H^2}|\phi|^2
\bigg\}\;,
\end{split}
\ee
where the canonical momenta $\pi_v$, $\pi_\phi$ are conjugate to the
variables  
$v$,
$\phi$, respectively, and 
\be
\label{hatk}
\hat{k}^2(t)=\frac{k^2}{f^2(t)}\;. 
\ee
For all fields in
the expression (\ref{Ham2}) one has  
$v(-{\bf k},t)=v^*({\bf k},t)$, etc. Note that the Hamiltonian
(\ref{Ham2}) explicitly depends on time via $\hat k$.
To simplify the Hamiltonian we perform a canonical transformation 
to variables $\tilde v$, $\tilde\pi_v$, etc.
The transformation is conveniently
parametrized by the generating function 
$\Phi(v,\phi,\tilde\pi_v,\tilde\pi_\phi; t)$ 
such that (see, e.g. \cite{Landau}),
\be
\tilde v=\frac{\d\Phi}{\d\tilde\pi_v^*}~,~~~
\pi_v=\frac{\d\Phi}{\d v^*}~,~~~\text{etc.}
\ee
The
Hamiltonian transforms in the following way,
\be
\label{Htrans}
{\cal H}_0\mapsto {\cal H}_0+\frac{\d \Phi}{\d t}\;,
\ee
where the partial time derivative acts on the explicit time dependence
of the generating function.
With the choice
\be
\label{simplify}
\begin{split}
\Phi=H\int d^3k\bigg\{-\frac{\hat k^2}{6H^2}|\tilde\pi_v|^2
+\left(\frac{\hat k^2}{6H^2}-\frac{1}{2}\right)
|\tilde\pi_\phi|^2\\
+\frac{\tilde\pi_v\tilde\pi_\phi^*}{2}
+\frac{\tilde\pi_v v^*}{\sqrt{6}}
-\frac{\tilde\pi_\phi v^*}{\sqrt{6}}
+\frac{\hat k^2}{\sqrt{6}H^2}\tilde\pi_\phi\phi^*
\bigg\}\;,
\end{split}
\ee
one obtains
\be
\label{Ham6}
\begin{split}
{\cal H}_0=\int& d^3k\bigg\{
\left(\hat k^2+\frac{3H^2}{4}\right)\frac{|\tilde\pi_v|^2}{2}
-H^2\tilde\pi_v\tilde\pi_\phi^*\\
&+\left(-\hat k^2+\frac{5H^2}{4}\right)
\frac{|\tilde\pi_\phi|^2}{2}
+\frac{|\tilde v|^2}{2}-\frac{|\tilde\phi|^2}{2}
\bigg\}\;.
\end{split}
\ee 
From this expression, it immediately follows that the mode
$\tilde\phi$ is ghost. 

Two comments are in order. First, the Hamiltonian (\ref{Ham6}) is
explicitly time-dependent via $\hat k^2(t)$ (see Eq.~(\ref{hatk})), 
and hence is not
conserved. Instead, the conserved energy in dS space-time
\cite{Abbott:1981ff,Deser:2001wx} is 
\be
\label{EdS}
E=\int d^3x~ T^0_\mu\,\bar\xi^\mu={\cal H}+\int d^3x~ T^0_i\,Hx_i\;,
\ee 
where $T^\nu_\mu$ is the energy-momentum tensor of the system, and
$\bar\xi^\mu=(-1, Hx_i)$  
is the Killing vector of dS space-time. 
However, when one considers
the system at distances much shorter than $1/H$ (which is the case
relevant to the ultraviolet 
stability), the second term in
(\ref{EdS}) can be neglected, and the energy coincides with the
Hamiltonian. 
Second, the expression (\ref{Ham6}) is not diagonal, reflecting the
resonance between the two modes. The situation again is similar
to the case of the Abelian model considered in Sec.\ref{abelian}. 
However, 
for the modes with wavelengths much smaller than the horizon size,
$\hat k^2\gg H^2$, one can neglect the
$O(H^2)$ terms in (\ref{Ham6}), and the two modes decouple. 

{\bf Acknowledgements} We are indebted to R.~Rattazzi for elucidating
discussions. We thank A.~Anisimov, F.~Bezrukov, C.~Deffayet,
S.~Dubovsky, R.~Maartens, V.~Rubakov, 
T.~Tanaka for useful conversations. We are
grateful to the organizers of the conference ``The Next Chapter in
Einstein's Legacy'' and the workshop ``Gravity and Cosmology'' at the
Yukawa Institute, Kyoto, at which this work began. S.S. is grateful
to the CERN Theory Division, where part of this work was done, for
hospitality. D.G  is grateful 
to Service de Physique Th\'{e}orique, Universit\'{e} Libre de Bruxelles, 
where this work was completed, for hospitality.    
The work of D.G. and S.S. has been supported in
part by the Russian Foundation for Basic Research grants 04-02-17448 and 
05-02-17363. The work of D.G. was also supported in part by the fellowship
of the "Dynasty" Foundation (awarded by the Scientific board of
ICFPM).  The work of S.S. was also supported in part by the grant of the
President of the Russian Federation MK-2205.2005.2. K.K. is supported
by PPARC.

\appendix
\section{Elimination of non-dynamical variables}
\label{appendix}
Substitution of the Fourier decomposition 
(\ref{Four}) into the action (\ref{effact}) yields
the Lagrangian for the helicity 0 sector:
\begin{widetext}
\be
\begin{split}
L_0=\int& d^3k \bigg\{-\frac{1}{2}|\dot{\alpha}_1-2H\alpha_1-2Hn|^2
-(\dot{\alpha}_1-2H\alpha_1-2Hn)
(\dot{\alpha}_2-2H\alpha_2-2|\hat k|\nu-2Hn)^*\\
&+(\dot{\phi}-H\phi)(\dot{\alpha}_1-2H\alpha_1-2Hn)^*
+\frac{1}{2}(\dot{\phi}-H\phi)
(\dot{\alpha}_2-2H\alpha_2-2|\hat k|\nu-2Hn)^*-\hat k^2\phi\alpha_1^*
+\frac{\hat k^2}{2}|\alpha_1|^2\\
&+H^2|\alpha_1|^2
+2H^2\alpha_1\alpha_2^*+H^2|\nu|^2
+n^*\left[\left(4H^2+2\hat k^2\right)\alpha_1+2H^2\alpha_2
-\hat k^2\phi\right]-\frac{9}{8}|\dot{\phi}|^2+\frac{9\hat k^2}{8}|\phi|^2
-\frac{9H^2}{2}|\phi|^2\bigg\}\;,
\end{split}
\label{Lagr1}
\ee
where $\hat k$ is defined by Eq. (\ref{hatk}).
In deriving
(\ref{Lagr1}) we omitted total derivative terms
and used the relations 
$(\alpha_1({\bf k}))^*=\alpha_1(-{\bf k})$, etc. 
The variable $\nu$ can be integrated out explicitly. From the
resulting Lagrangian one  
determines the canonical momenta:
\bseq
\label{momenta}
\begin{align}
&\pi_1\equiv\frac{\d L}{\d\dot{\alpha}_1^*}
=-\left(1+\frac{2\hat{k}^2}{H^2}\right)
\left(\dot{\alpha_1}-\frac{3H}{2}\alpha_1-2Hn\right)
-\left(\dot{\alpha_2}-\frac{3H}{2}\alpha_2-2Hn\right)
+\left(1+\frac{\hat{k}^2}{H^2}\right)\left(\dot{\phi}-\frac{5H}{2}\phi\right)
\label{pi1}\;,\\
&\pi_2\equiv\frac{\d L}{\d\dot{\alpha}_2^*}
=-\left(\dot{\alpha_1}-\frac{3H}{2}\alpha_1-2Hn\right)
+\frac{1}{2}\left(\dot{\phi}-\frac{5H}{2}\phi\right)
\label{pi2}\;,\\
&\pi_{\phi}\equiv\frac{\d L}{\d\dot{\phi}^*}
=\left(1+\frac{\hat{k}^2}{H^2}\right)
\left(\dot{\alpha_1}-\frac{3H}{2}\alpha_1-2Hn\right)
+\frac{1}{2}\left(\dot{\alpha_2}-\frac{3H}{2}\alpha_2-2Hn\right)
-\left(\frac{9}{4}+\frac{\hat{k}^2}{2H^2}\right)
\left(\dot{\phi}-\frac{5H}{2}\phi\right)\;,
\label{pi}
\end{align}
\eseq
and the Hamiltonian:
\[
%\be
%\label{Ham1}
\begin{split}
&{\cal H}_0=\int  d^3k\bigg\{
-\frac{1}{12}|\pi_1|^2+\left(\frac{5}{12}+\frac{\hat{k}^2}{H^2}\right)|\pi_2|^2
-\frac{1}{3}|\pi_{\phi}|^2
-\frac{7}{6}\pi_1\pi_2^*-\frac{1}{3}\pi_1\pi_{\phi}^*
-\frac{1}{3}\pi_2\pi_{\phi}^*
+\frac{3H}{2}\pi_1\alpha_1^*+\frac{3H}{2}\pi_2\alpha_2^*
+\frac{5H}{2}\pi_{\phi}\phi^*\\
&+\hat{k}^2\alpha_1\phi^*-\left(\frac{\hat{k}^2}{2}+H^2\right)|\alpha_1|^2
-2H^2\alpha_1\alpha_2^*-\frac{9}{8}\hat{k}^2|\phi|^2
+n^*\big[2H\pi_1+2H\pi_2-(4H^2+2\hat{k}^2)\alpha_1
-2H^2\alpha_2+\hat{k}^2\phi\big]
\bigg\}. 
\end{split}
%\ee
\]
\end{widetext}
Let us study the set of constraints of the system.
From the Hamiltonian one obtains the primary constraint
\be
\label{constr1}
\chi_1\equiv\pi_1+\pi_2-\frac{\hat{k}^2+2H^2}{H}\alpha_1-H\alpha_2
+\frac{\hat{k}^2}{2H}\phi=0\;.
\ee
Taking the time
derivative of Eq.~(\ref{constr1}), using equations of motion
and the relation
\be
\frac{d\hat{k}^2}{dt}=-2H\hat{k}^2,
\ee 
we find the secondary constraint,
\be
\label{constr2}
\chi_2\equiv\frac{\pi_1}{H}+
\left(\frac{1}{H}-\frac{2\hat{k}^2}{3H^3}\right)\pi_2
+\frac{2}{3H}\pi_\phi-\frac{4}{3}\alpha_1-\frac{2}{3}\alpha_2=0\;.
\ee
Finally, vanishing of the time derivative of Eq.~(\ref{constr2})
implies the constraint
\be
\label{constr3}
2n+2\alpha_1+\alpha_2=0\;.
\ee
Note, that Eq.~(\ref{constr3}) is equivalent to the tracelessness
condition $A_{\mu}^\mu=0$. We use Eq.~(\ref{constr3}) 
to eliminate the variable $n$ from the Hamiltonian. 
The constraints (\ref{constr1}), (\ref{constr2})
are
second class 
and have canonical Poisson bracket: 
\be
\label{Poisson}
\{\chi_1,\chi_2\}=1\;.  
\ee 
Let us reduce the number of degrees of
freedom by performing a canonical transformation to variables
$\tilde\pi_1$, $\tilde\alpha_1$, etc., such that 
\be
\label{req}
\tilde\pi_2=\chi_1~,~~~~\tilde\alpha_2=\chi_2\;.
\ee
To find a suitable transformation we introduce the generating function
$\Phi$ depending on the old coordinates $\alpha_1, \alpha_2, \phi$ and
the new momenta $\tilde\pi_1, \tilde\pi_2, \tilde\pi_\phi$ such that,
\be
\tilde\alpha_1=\frac{\d\Phi}{\d\tilde\pi_1^*}~,~~~
\pi_1=\frac{\d\Phi}{\d\alpha_1^*}~,~~~\text{etc.}
\ee
The requirements (\ref{req}) imply 
the following equations:
\begin{align}
\tilde\pi_2=&\frac{\d\Phi}{\d\alpha_1^*}+\frac{\d\Phi}{\d\alpha_2^*}
-\frac{\hat{k}^2+2H^2}{H}\alpha_1-H\alpha_2
+\frac{\hat{k}^2}{2H}\phi\;,\notag\\
\frac{\d\Phi}{\d\tilde\pi_2^*}&=\frac{1}{H}\frac{\d\Phi}{\d\alpha_1^*}
+\left(\frac{1}{H}-\frac{2\hat{k}^2}{3H^3}\right)\frac{\d\Phi}{\d\alpha_2^*}
\notag\\
&~~~~
+\frac{2}{3H}\frac{\d\Phi}{\d\phi^*}
-\frac{4}{3}\alpha_1-\frac{2}{3}\alpha_2
\notag\;.
\end{align}
As a solution we choose
\[
%\be
%\label{Phi1}
\begin{split}
\Phi&=\int d^3 k\bigg\{\frac{2\hat{k}^2}{3H^3}\tilde\pi_1\tilde\pi_2^*
+\left(\frac{1}{2H}-\frac{\hat{k}^2}{9H^3}\right)|\tilde\pi_2|^2
+\frac{2}{3H}\tilde\pi_2\tilde\pi_\phi^*\\
&+\tilde\pi_1\alpha_1^*-\tilde\pi_1\alpha_2^*
+\frac{2}{3}\tilde\pi_2\alpha_1^*+\frac{1}{3}\tilde\pi_2\alpha_2^*
+\tilde\pi_\phi\phi^*\\
&+\frac{\hat{k}^2+H^2}{2H}|\alpha_1|^2+H\alpha_1\alpha_2^*
-\frac{\hat{k}^2}{2H}\alpha_1\phi^*+\frac{3\hat{k}^2}{8H}|\phi|^2
\bigg\}\;.
\end{split}
%\ee
\]
%\vskip -0.4cm
The transformation generated by this function has the form,
\bseq
\label{can1}
\begin{align}
\tilde\alpha_1
=&\frac{2\hat k^2}{3H^3}\tilde\pi_2+\alpha_1-\alpha_2\;,\\
\tilde\alpha_2
=&\frac{2\hat k^2}{3H^3}\tilde\pi_1
+\left(\frac{1}{H}-\frac{2\hat k^2}{9H^3}\right)\tilde\pi_2
+\frac{2}{3H}\tilde\pi_\phi\notag\\
&+\frac{2}{3}\alpha_1+\frac{1}{3}\alpha_2\;,\\
\tilde\phi
=&\frac{2}{3H}\tilde\pi_2+\phi\;,\\
\pi_1
=&\tilde\pi_1+\frac{2}{3}\tilde\pi_2
+\frac{\hat k^2+H^2}{H}\alpha_1+H\alpha_2
-\frac{\hat k^2}{2H}\phi\;,\\
\pi_2
=&-\tilde\pi_1+\frac{1}{3}\tilde\pi_2+H\alpha_1\;,\\
\pi_\phi
=&\tilde\pi_\phi-\frac{\hat k^2}{2H}\alpha_1
+\frac{3\hat k^2}{4H}\phi \;.
\end{align}
\eseq
Using these expressions one can check 
explicitly that the conditions (\ref{req})
are satisfied. Making use of the constraints we set 
$\tilde\pi_2=\tilde\alpha_2=0$,
and we are left with two dynamical variables 
$v\equiv\tilde\alpha_1=\alpha_1-\alpha_2$, $\tilde\phi=\phi$ and
their conjugate momenta. In the derivation of the resulting 
Hamiltonian one should account for the explicit time dependence of the
generating function according to Eq.~(\ref{Htrans}). 
A straightforward calculation yields the Hamiltonian 
(\ref{Ham2}) of the main text.

%%%%%%%%%%%%%%%%%%%%%%%%%%%%%%%%%%%%%%%%%%%%%%%%%%%%%%%%%%%%%%%%%%%%%%%%%%%%%


\begin{thebibliography}{99}
%\cite{Riess:1998cb}
\bibitem{Riess:1998cb}
  A.~G.~Riess {\it et al.}  [Supernova Search Team Collaboration],
  %``Observational Evidence from Supernovae for an Accelerating Universe and a
  %Cosmological Constant,'' 
  Astron.\ J.\  {\bf 116}, 1009 (1998)
  [arXiv:astro-ph/9805201];
  %%CITATION = ASTRO-PH 9805201;%%

%\cite{Perlmutter:1998np}
%\bibitem{Perlmutter:1998np}
  S.~Perlmutter {\it et al.}  [Supernova Cosmology Project Collaboration],
  %``Measurements of Omega and Lambda from 42 High-Redshift Supernovae,''
  Astrophys.\ J.\  {\bf 517}, 565 (1999)
  [arXiv:astro-ph/9812133];
  %%CITATION = ASTRO-PH 9812133;%%

%\cite{Hinshaw:2003ex}
%\bibitem{Hinshaw:2003ex}
  G.~Hinshaw {\it et al.},
  %``First Year Wilkinson Microwave Anisotropy Probe (WMAP) Observations:
  %Angular Power Spectrum,''
  Astrophys.\ J.\ Suppl.\  {\bf 148}, 135 (2003)
  [arXiv:astro-ph/0302217].
  %%CITATION = ASTRO-PH 0302217;%%

%\cite{Cline:2003gs}
\bibitem{Cline:2003gs}
  J.~M.~Cline, S.~Jeon and G.~D.~Moore,
  %``The phantom menaced: Constraints on low-energy effective ghosts,''
  Phys.\ Rev.\ D {\bf 70}, 043543 (2004)
  [arXiv:hep-ph/0311312].
  %%CITATION = HEP-PH 0311312;%%


%\cite{Dvali:2000hr}
\bibitem{Dvali:2000hr}
  G.~R.~Dvali, G.~Gabadadze and M.~Porrati,
  %``4D gravity on a brane in 5D Minkowski space,''
  Phys.\ Lett.\ B {\bf 485} (2000) 208
  [arXiv:hep-th/0005016].
  %%CITATION = HEP-TH 0005016;%%

%\cite{Deffayet:2001uk}
\bibitem{Deffayet:2001uk}
  C.~Deffayet, G.~R.~Dvali, G.~Gabadadze and A.~I.~Vainshtein,
  %``Nonperturbative continuity in graviton mass versus perturbative
  %discontinuity,''
  Phys.\ Rev.\ D {\bf 65} (2002) 044026
  [arXiv:hep-th/0106001].
  %%CITATION = HEP-TH 0106001;%%

%\cite{Nicolis:2004qq}
\bibitem{Nicolis:2004qq}
  A.~Nicolis and R.~Rattazzi,
  %``Classical and quantum consistency of the DGP model,''
  JHEP {\bf 0406}, 059 (2004)
  [arXiv:hep-th/0404159].
  %%CITATION = HEP-TH 0404159;%%

%\cite{Deffayet:2001pu}
\bibitem{Deffayet:2001pu}
  C.~Deffayet, G.~R.~Dvali and G.~Gabadadze,
  %``Accelerated Universe from gravity leaking to extra dimensions,''
  Phys.\ Rev.\ D {\bf 65}, 044023 (2002)
  [arXiv:astro-ph/0105068].
  %%CITATION = ASTRO-PH 0105068;%%

%\cite{Luty:2003vm}
\bibitem{Luty:2003vm}
  M.~A.~Luty, M.~Porrati and R.~Rattazzi,
  %``Strong interactions and stability in the DGP model,''
  JHEP {\bf 0309}, 029 (2003)
  [arXiv:hep-th/0303116].
  %%CITATION = HEP-TH 0303116;%%

%\cite{Koyama:2005tx}
\bibitem{Koyama:2005tx}
  K.~Koyama,
  %``Ghosts in the self-accelerating brane universe,''
  Phys.\ Rev.\ D, in press [arXiv:hep-th/0503191].
  %%CITATION = HEP-TH 0503191;%%

%\cite{Abbott:1981ff}
\bibitem{Abbott:1981ff}
  L.~F.~Abbott and S.~Deser,
  %``Stability Of Gravity With A Cosmological Constant,''
  Nucl.\ Phys.\ B {\bf 195}, 76 (1982).
  %%CITATION = NUPHA,B195,76;%%

%\cite{Deser:2001wx}
\bibitem{Deser:2001wx}
  S.~Deser and A.~Waldron,
  %``Stability of massive cosmological gravitons,''
  Phys.\ Lett.\ B {\bf 508}, 347 (2001)
  [arXiv:hep-th/0103255].
  %%CITATION = HEP-TH 0103255;%%

\bibitem{Arnold} V.I.~Arnold, ``Mathematical Methods of Classical
Mechanics'', Springer-Verlag, NY, 1978.

%\cite{Landau}
\bibitem{Landau}
  L.~D.~Landau and E.~M.~Lifshitz,
  Mechanics.- Oxford: Pergamon, 1976.- (Course Theor. Phys., v.1)

\end{thebibliography}
\end{document}